\documentclass[prb,twocolumn,preprintnumbers,superscriptaddress]{revtex4}

\usepackage{graphicx,epstopdf}% Include figure files
\usepackage{amssymb,amsmath}
\usepackage{dcolumn}% Align table columns on decimal point
\usepackage{bm}% bold math
\usepackage{color}
\usepackage{enumerate}
\usepackage{hyperref}

\usepackage{soul}

\begin{document}

\title{Exciton-exciton interaction in transition-metal dichalcogenide monolayers}

\author{V. Shahnazaryan}
\affiliation{Science Institute, University of Iceland IS-107, Reykjavik, Iceland}
\affiliation{ITMO University, St. Petersburg 197101, Russia}
\affiliation{Russian-Armenian (Slavonic) University, Yerevan 0051, Armenia}

\author{I. Iorsh}
\affiliation{ITMO University, St. Petersburg 197101, Russia}

\author{I. A. Shelykh}
\affiliation{Science Institute, University of Iceland IS-107, Reykjavik, Iceland}
\affiliation{ITMO University, St. Petersburg 197101, Russia}

\author{O. Kyriienko}
\affiliation{The Niels Bohr Institute, University of Copenhagen, Copenhagen DK-2100, Denmark}

\begin{abstract}
We study theoretically the Coulomb interaction between excitons in transition metal dichalcogenide (TMD) monolayers. We calculate direct and exchange interaction for both ground and excited states of excitons. The screening of the Coulomb interaction, specific to monolayer structures, leads to the unique behavior of the exciton-exciton scattering for excited states, characterized by the non-monotonic dependence of the interaction as function of the transferred momentum. We find that the nontrivial screening enables the description of TMD exciton interaction strength by approximate formula which includes exciton binding parameters. The influence of screening and dielectric environment on the exciton-exciton interaction was studied, showing qualitatively different behavior for ground state and excited states of excitons. Furthermore, we consider exciton-electron interaction, which for the excited states is governed by the dominant attractive contribution of the exchange component, which increases with the excitation number. The results provide a quantitative description of the exciton-exciton and exciton-electron scattering in transition metal dichalcogenides, and are of interest for the design of  perspective nonlinear optical devices based on TMD monolayers.
\end{abstract}

\maketitle

\section{Introduction}

The physics of excitons and associated optical phenomena was greatly influenced by recent discoveries in the domain of transition metal dichalcogenide (TMD) materials [\onlinecite{Novoselov2016}]. They can both exist in the bulk and monolayer configurations and possess a set of peculiar properties which differ them from other semiconductor materials. TMD monolayers are of particular interest in this context --- they are atomically thin, are characterized by the direct bandgap favorable for optical interband transitions and exciton formation and demonstrate peculiar interplay of spin and valley effects. All this give them certain advantages for the use in optoelectronics as compared to semimetallic graphene.

Extensive studies of excitonic properties of TMD monolayers started immediately after their discovery [\onlinecite{Mak2010}]. In striking contrast to bulk and quasi-2D structures, the different screening in 2D monolayer governs the deviation of the interparticle Coulomb interaction from the standard form, and ultimately leads to unusual properties of the excitons in TMD structures [\onlinecite{Keldysh,Cudazzo2011}].
The exciton binding energies and absorption spectra in various TMD monolayers were measured experimentally [\onlinecite{Zhao,Shan,Zhu,Ceballos,Shang,Schmidt}] and calculated from the first principles [\onlinecite{Komsa1,Komsa2,Yakobson,Qiu1,Qiu2,Rama,Berkelbach,Lambrecht}]. The results have shown huge increase of the exciton binding energy (up to 1~eV) [\onlinecite{Rasmussen2015}], as compared to conventional semiconductors, and the non-hydrogenic behavior of the excitonic series [\onlinecite{ChernikovPRL}].
Further investigations cover measurements of exciton lifetimes and linewidths in monolayers [\onlinecite{Schaibley,Robert,Selig}], as well as electric field control of the excitonic properties [\onlinecite{Heinz,Scharf}]. Moreover, the rich many-body physics in TMD materials was confirmed by observation of more complex particles, such as trions and biexcitons [\onlinecite{You,Zhang,Sie,Plechinger}] as well as interlayer excitons in bilayer structures [\onlinecite{Rivera,Bellus,Butov}]. Additionally, the hybrid exciton-electron systems in TMDs were considered [\onlinecite{Sidler2017}].
%%%
\begin{figure}
    \includegraphics[width=0.8\linewidth]{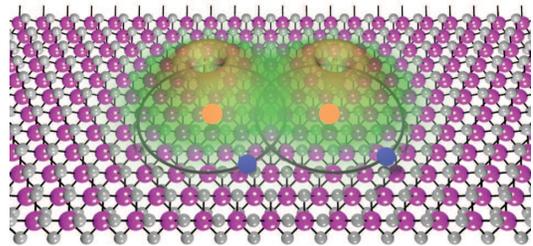}
    \caption{Sketch of the system. A transition metal dichalcogenide monolayer hosts excitonic quasiparticles formed by electrons (blue circles) and holes (red circles). The scattering between two excitons corresponds to the Coulomb interaction between carriers, consisting of the direct and exchange contributions. The latter is dependent on the exciton wavefunction overlap, shown in green.}
    \label{fig:sketch}
\end{figure}
%%%

Excellent optical properties of TMD monolayers put them as a prominent platform for optoelectronical applications. For instance, the large binding energy of excitons allowed to study excitonic physics at elevated temperatures, and observe the excitons with high principal quantum numbers [\onlinecite{ChernikovPRL}]. The large oscillator strength allows to couple excitons strongly to an optical microcavity mode, and study strong light-matter coupling at room temperature [\onlinecite{Dufferwiel,Lundt1,Lundt2}]. The particular spin-orbit interaction for the bands leads to non-trivial valley dynamics and spin properties, also suggested to be potentially interesting for the quantum information processing [\onlinecite{Wang2016}]. Finally, $\chi_2$ nonlinear response of TMD monolayers was predicted, making it suitable for the observation of the nonlinear quantum optical effects [\onlinecite{Glazov2017}]. %
There are several experimental investigations of TMD monolayer properties in the strong excitation regime, manifesting itself in various intriguing phenomena, including spectral peak broadening [\onlinecite{Sim}], exciton-exciton annihilation [\onlinecite{Kumar}], and giant bandgap renormalization (up to 500 meV) in the vicinity of Mott transition [\onlinecite{ChernikovNat,Aivazian}]. However, to the best of our knowledge, theoretical investigations of the interexciton interactions in TMD monolayers are lacking so far.
%
%Various analytical approaches were proposed to study the exciton binding energy, oscillator strength, optical transitions and spin properties [\onlinecite{Berghauser,GlazovTMD,Wu,Li}].

Motivated by the aforementioned advances, we consider the nonlinear properties of excitons in a TMD monolayer. The system is well-suitable for the observation of highly excited states of excitons, and similarly to bulk semiconductors [\onlinecite{Kazimierczuk2014}], can allow for studying nonlinear interaction between Rydberg excitons. In the paper, we calculate the exciton-exciton interaction in TMD structures, considering both ground and excited states of excitons. We find that the interaction of excited states exhibits non-monotonic dependence on the exchanged momentum and is attractive. We provide the analytical formula to quantitatively estimate the maximal exciton-exciton interaction strength, which differs from those for the III-V group semiconductors. Finally, we calculate the exciton-electron matrix elements of scattering for both direct and exchange terms.

\section{Excitonic spectrum in TMD monolayer}

To study the interparticle interactions in TMD monolayers, one should take into account structural peculiarities of such materials. Namely, the atomic thickness of the layer and discontinuity of the dielectric screening on the monolayer interface modifies the Coulomb interaction to the following form [\onlinecite{Keldysh}]:
\begin{equation}
V(r)=\frac{e_1 e_2}{4 \pi \varepsilon_0} \frac{\pi }{2 r_0}\left[H_0\left(\frac{r}{r_0}\right)-Y_0\left(\frac{r}{r_0}\right)\right],
\label{pot_0}
\end{equation}
where $e_1$, $e_2$ denote the charge of particles, $r$ is the interparticle distance, and $r_0$ is a quantity describing the polarizability of the monolayer. $H_0$ and $Y_0$ are zeros order Struve and Bessel functions of the first kind, respectively. The modification of Coulomb interaction results in the qualitative change of the excitonic spectrum [\onlinecite{Berkelbach,ChernikovPRL}], which in this case cannot be considered as common 2D hydrogenic spectrum of the form $E_n=\mu e^4 /[2(4 \pi \varepsilon_0\varepsilon)^2\hbar^2(n-1/2)^2]$, where $n$ is a principal quantum number of the exciton, $\mu$ is reduced mass of an electron- hole pair, $\varepsilon$ corresponds to the static dielectric screening constant, and $\varepsilon_0$ is the vacuum permittivity.

The excitonic states should be found as eigenstates of the Hamiltonian
\begin{equation}
\hat{H}_{\mathrm{exc}}=-\frac{\hbar^2}{2\mu}\Delta+V(r),
\label{ExcHam}
\end{equation}
with $V(r)$ is taken in the form of Eq. (\ref{pot_0}). As first approximation one can use variational method, where the trial functions are similar to the conventional 2D excitonic functions and excitonic Bohr radius plays a role of variational parameter [\onlinecite{Portnoi}]:
\begin{align}
\label{psi_n}
&\psi_{n,m}(r)=\frac{1}{\sqrt{2}\lambda_{n}}\sqrt{\frac{(n-|m|-1)!}{(n+|m|-1)!(n-1/2)^3}} \\ \notag
&\hspace{10mm} \left(\frac{r}{(n-1/2)\lambda_{n}}\right)^m  \exp\left[-\frac{r}{(2n-1)\lambda_{n}}\right] \\ \notag
&\hspace{10mm} L_{n-|m|-1}^{2|m|}\left[\frac{r}{(n-1/2)\lambda_{n}}\right] \frac{1}{\sqrt{2\pi}} e^{im\varphi}.
\end{align}
Here $L_n^m[x]$ denotes associated Laguerre polynomial, $\lambda_{n}$ is a variational parameter, and $m$ is an angular momentum quantum number. Contrary to the conventional quantum well exciton, where all states have the same radial characteristic --- two-dimensional Bohr radius, in the case of a monolayer the spatial parameter $\lambda_n$ changes from state to state.

To be specific, we consider WS$_2$ monolayer, noting however that all results are of general character and are applicable for the whole family of TMD monolayers. Accurate calculation of exciton series confirmed by experimental data was done in Ref. [\onlinecite{ChernikovPRL}], where the value of polarizibility parameter $r_0$ was found to be equal to $7.5$~nm. Here, we reproduce these results by the binding energy minimization using $\lambda_n$ as a variational parameter. The corresponding values of the exciton energies and spatial characteristics $\lambda_n$ are presented in the Table 1. Note that while the energies of the lower states are essentially non-hydrogenic, for the states starting from $n=3$ the conventional $n^{-2}$ energy dependence can be observed. Correspondingly, the saturation of the $\lambda_n$ values can be seen for higher states.
%%%
\begin{table}
\centering
    \begin{tabular}{|c|c|c|}
        \hline
        n & $\lambda_{n}~(\mathrm{nm})$ & $E_{n}$ (meV)  \\ \hline
        1 & 1.7                         & 320            \\ \hline
        2 & 0.65                        & 160            \\ \hline
        3 & 0.45                        & 90             \\ \hline
        4 & 0.35                        & 60             \\ \hline
        5 & 0.3                         & 50             \\ \hline
    \end{tabular}
    \caption{Spatial characteristics ($\lambda_n$) and energies of excitons ($E_n$) of different states $n$ calculated for the WS$_2$ monolayer.}
\end{table}
%%%

\section{Exciton-exciton interaction}

Analyzing the asymptotic behavior of the potential given by Eq. (\ref{pot_0}) one can find its accurate approximate expression [\onlinecite{Cudazzo2011}]
\begin{equation}
V(r)= - \frac{e_1 e_2}{4 \pi \varepsilon_0} \frac{1}{r_0}\left[\mathrm{ln}\left(\frac{r}{r+r_0}\right)-(\gamma-\mathrm{ln}2)e^{-\frac{r}{r_0}} \right],
\label{pot}
\end{equation}
which is used in further calculations. $\gamma$ denotes Euler gamma constant. To calculate interactions between TMD monolayer excitons in the ground and excited states, we employ the method similar to those used by us before for the case of III-V semiconductor quantum well structures [\onlinecite{Shahnazaryan2016}]. It represents the generalization of the Coulomb scattering formalism for the ground state excitons in quantum wells developed in the Ref. [\onlinecite{Tassone1999,Ciuti1998}]. The wavefunction of an exciton with a wave vector $\textbf{Q}$ can be written in the form
\begin{equation}
\Psi_{\mathbf{Q},n,m}(\mathbf{r}_e,\mathbf{r}_h)=\frac{1}{\sqrt{A}}\exp[i\mathbf{Q}(\beta_e\mathbf{r}_e+\beta_h\mathbf{r}_h)]\psi_{n,m}(|\mathbf{r}_e-\mathbf{r}_h|),
\label{Psi}
\end{equation}
where $\mathbf{r}_e, \mathbf{r}_h$ are the radius vectors of an electron and a hole, respectively, $A$ denotes the normalization area. The coefficients $\beta_e, \beta_h$ are defined as $\beta_{e(h)}=m_{e(h)}/(m_e+m_h)$, where $m_{e(h)}$ is the mass of an electron (hole). The wavefunction of relative motion of an electron and a hole motion is described by Eq. (\ref{psi_n}).
%%%
\begin{figure}[t]
    \includegraphics[width=1.0\linewidth]{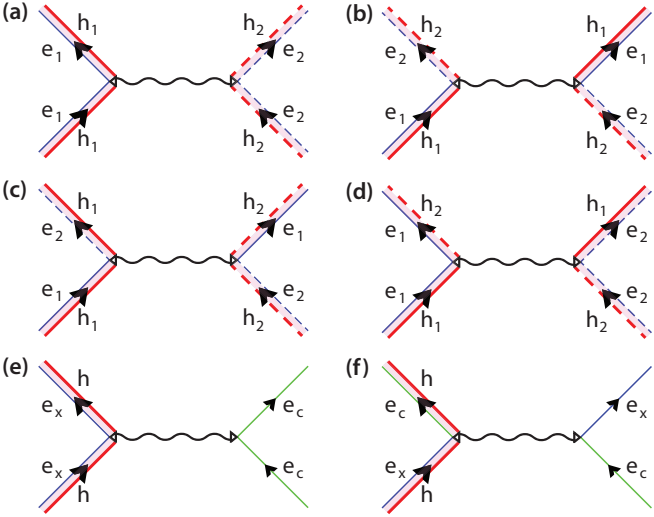}
    \caption{The schematic representation of the exciton-exciton scattering \textbf{(a)-(d)}. Panels correspond to: (a) direct, (b) exciton exchange, (c) electron exchange, and (d) hole exchange interactions. Blue and red solid lines denote an electron ($e_1$) and a hole ($h_1$) of the first exciton exciton, and the dashed lines correspond to an electron and a hole of the second exciton marked with $e_2$ and $h_2$. The exciton-electron scattering diagrams are shown in panels \textbf{(e)-(f)}, describing the direct (e) and exchange (f) interaction. Green solid line ($e_c$) denotes a free electron.}
    \label{fig:scheme}
\end{figure}
%%%

We consider the interaction of the excitons in the same states with parallel spin projections. In this case the process of Coulomb scattering in reciprocal space with transfer of wave vector $\mathbf{q}$ can be presented in the form
\begin{equation}
(n,m,\mathbf{Q}) + (n,m,\mathbf{Q}') \rightarrow (n,m,\mathbf{Q}+\mathbf{q})+(n,m,\mathbf{Q}'-\mathbf{q}),
\label{scattering}
\end{equation}
which can be represented graphically by scattering diagrams in Fig. \ref{fig:scheme}.
Using the wave function symmetrization procedure the total interaction may be presented as linear combination of the interaction channels, including direct interaction, and electron, hole, exciton exchange terms, as schematically depicted in Fig. \ref{fig:scheme} (a)-(d). It was shown previously [\onlinecite{Ciuti1998,Shahnazaryan2016,Tassone1999}] that in the wide region of exchanged wave vectors $q\leq 1/\lambda_1$ the interaction of excitons is determined by the exchange terms, while the direct interaction in negligibly small. The latter becomes dominant for large values of $\textbf{q}$, governing the long range behavior of the interaction. Assuming the initial wave vectors being equal and setting $\textbf{Q}=\textbf{Q}'=0$, for the total interaction we have (see Appendix A for the details and definitions):
\begin{align}
\label{Htot}
&V_{\mathrm{tot}}(n,m,\mathbf{q})=\frac{e^2}{4 \pi \varepsilon_0}\frac{\lambda_{1}}{A} I_{\mathrm{tot}}(n,m,q\lambda_1),\\
&I_{\mathrm{tot}}(n,m,q\lambda_1) =I_{\mathrm{dir}}(n,m,q\lambda_1)+I^X_{\mathrm{exch}}(n,m,q\lambda_1)  \notag \\
&+I^e_{\mathrm{exch}}(n,m,q\lambda_1)+I^h_{\mathrm{exch}}(n,m,q\lambda_1) \approx 2 I^e_{\mathrm{exch}}(n,m,q\lambda_1),
\label{Itot}
\end{align}
where indices $e$, $h$, $X$ stand for the electron, hole, and exciton exchange integrals, respectively.
In principle, the interaction processes between excitons with different spin projections can be accounted for. However, they involve spin-flip processes, and typically contain only direct interaction channel [\onlinecite{Kyriienko2012}].
%%%
\begin{figure}[t]
    \includegraphics[width=1.0\linewidth]{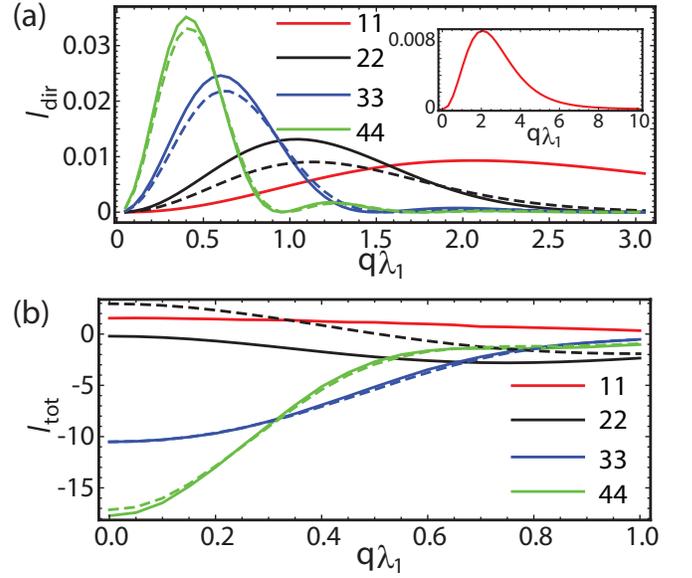}
    \caption{Dimensionless integrals corresponding to the direct interaction matrix element (a), and total interaction energy  (b) of excitons in TMD monolayer, plotted as function of the transferred wave vector. Solid lines correspond to $s$ states ($m=0$) and dashed lines denote $p$ states ($m=1$). The interaction has similar form for both type excitons, being attractive for the excited states. The difference appears for $n=2$ state, where for $s$ state there is attraction with the absolute maxima at intermediate momenta, and for $p$ state there are attraction and repulsion regions.}
    \label{exc}
\end{figure}
%%%

We calculate direct and total interaction as a function of the scattered momentum exploiting the multidimensional Monte-Carlo integration [\onlinecite{Hahn_VEGAS}]. The results of the calculation are shown in Fig. \ref{exc}.
The direct interaction as function of the exchanged momentum is repulsive, and its peak-shaped dependence becomes narrower with increase of the principal quantum number of the scattered excitons. The total interaction is fully governed by the exchange term, which is non-zero at $q \lambda_1 \rightarrow 0$. It is repulsive for the ground state and attractive for the excited states. This behavior is qualitatively similar to quantum well exciton interaction [\onlinecite{Shahnazaryan2016}].
However, the screened nature of Coulomb interaction imposes peculiarities in the TMD exciton-exciton  interaction behavior. Namely, the crucial difference of the monolayer exciton interaction appears in the dependence of $2s$ state interactions, which demonstrate potential minima for the nonzero exchange momenta $q$. It should be noted that the interaction of $2p$ excitons demonstrates similar properties, being repulsive for zero exchange momenta and having attraction peak at intermediate momenta. This non-monotonic behavior can be expected to lead to different condensation processes for TMD polaritons [\onlinecite{Matuszewski2012}].
%%%
\begin{figure}[t]
    \includegraphics[width=1.0\linewidth]{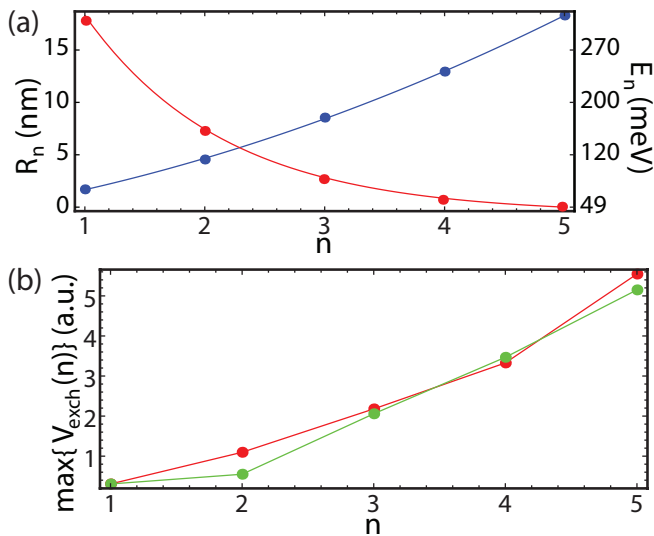}
    \caption{ (a) The dependence of exciton radius (blue curve) and energy (red curve) on the principal quantum number. (b) Exchange interaction energy as a function of the principal quantum number of a TMD exciton. The green curve corresponds to the numerical calculation, and the red line shows the qualitative estimate using Eq. (\ref{Uexch_simp}).}
    \label{fig:scaling}
\end{figure}
%%%

Next, we search for the compact analytical formula to describe the exciton-exciton interaction in TMD monolayers, considering both ground and excite states scattering. Previously it was shown that the exchange interaction of GaAs quantum well ground state excitons can be described by the formula $V_{\mathrm{exch}}^{\mathrm{QW}}= {6 E_b a_B^2}/A$, where $a_B$ and $E_b$ denote Bohr radius and binding energy of quantum well exciton, respectively [\onlinecite{Tassone1999}]. The numerical prefactor 6 comes from the calculation of exchange integrals.

Following the analogy, we search for similar dependence for the exciton series in TMD monolayer. Fig.~\ref{fig:scaling}(a) presents the dependence of the radius and energy of exciton states on their principal quantum number. While the radius increases quadratically (as in the case of the conventional Rydberg series), the energy dependence for the first few states drops superpolynomially with $n$. The latter allows us to approximate the exchange interaction dependence by the formula
\begin{equation}
\label{Uexch_simp}
V_{\mathrm{exch}}(n)={\alpha E_n R_n^2}/A,
\end{equation}
where $R_n$ and $E_n$ denote the radius and energy of $n$-th exciton state, respectively, and $\alpha$ is a fitting constant. The green line in Fig. \ref{fig:scaling}(b) denotes the dependence of the exchange interaction strength on the principal quantum number, unveiling close-to-linear dependence starting from the $n=2$ state. The red curve shows the estimate by Eq. (\ref{Uexch_simp}), where we chose the parameter $\alpha=2.07$, which gives the exact fitting for the ground state. It is worth mentioning that despite the smaller pre-factor, for the case of the quantum well with the similar material parameters the interaction would be weaker. Namely, taking the reduced effective mass characteristic to WS$_2$ monolayer, $\mu = 0.16 m_0$ [\onlinecite{ChernikovPRL}], the interaction strength between ground state excitons in a quantum well can be estimated as $V_{\mathrm{exch}}^{\mathrm{QW}}= 6 \frac{\hbar^2}{2\mu A}$. Comparing it to the TMD estimate $V_{\mathrm{exch}}^{\mathrm{TMD}}= 2.07 E_{1s} \lambda_{1s}^2/A$, where $E_{1s}$ and $\lambda_{1s}$ values are taken from Table 1, we get the ratio
\begin{equation}
V_{\mathrm{exch}}^{\mathrm{TMD}} / V_{\mathrm{exch}}^{\mathrm{QW}} = 1.34.
\label{ratio}
\end{equation}

The reason beyond this is peculiar interaction screening in TMD monolayers, leading to the exciton effective radius value larger than for the conventional Coulomb potential.

It should be noted that the close agreement between the exact calculation of interaction and its qualitative estimate is possible only because of the rapid decrease of the exciton energy for the lower excitonic states in TMD monolayer. On the contrary, in semiconductor heterostructures the energy drops quadratically, $E_n \sim n^{-2}$, obeying Rydberg rule. The corresponding estimate thus predicts quadratic growth of the interaction strength. However the exact calculation shows the linear dependence of the exchange term on quantum number for quantum well [\onlinecite{Shahnazaryan2016}], meaning that the estimate is not reasonable for that case.
%%%
\begin{figure}[t]
    \includegraphics[width=1.0\linewidth]{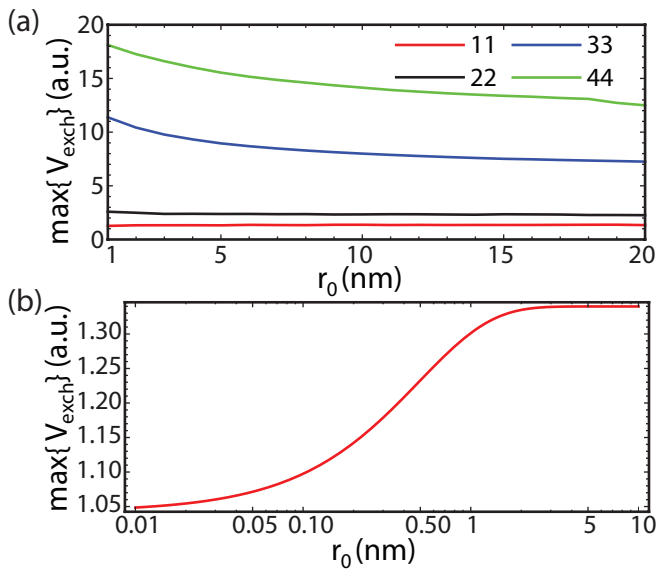}
    \caption{(a) Maximum of exchange interaction energy plotted as a function of the screening length $r_0$. Interaction between different excitonic states is considered. (b) Maximum of exchange interaction energy of ground excitonic states plotted as a function of the screening length $r_0$. Here, $r_0$ axis is extended to show small screening length limit, and we use the logarithmic scale. }
    \label{fig:screening}
\end{figure}
%%%

Finally, we proceed with the discussion of the influence of screening on the interexciton interaction. While previously we focused on a structure of particular configuration, discussed in Ref. [\onlinecite{ChernikovPRL}], the obtained results are expected to be qualitatively valid for other configurations as well. For instance, an additional factor is the presence of a substrate, which can substantially modify the optical properties of a sample. Particularly, the strong modulation of monolayer bandgap by the dielectric environment was studied in Ref. [\onlinecite{Ryou}]. In our model the screening length $r_0$ is influenced by the dielectric permittivity of a monolayer ($\varepsilon$) and substrate ($\varepsilon_{1,2}$) as $r_0=d \varepsilon / (\varepsilon_1+\varepsilon_2)$, where $d$ denotes the thickness of a monolayer, and $\varepsilon_{1,2}$ stand for dielectric permittivity of substrate and cover layer, respectively [\onlinecite{Keldysh, Berkelbach}]. Here we vary the screening length in a wide range and study it influence on the exciton-exciton interaction strength. In Fig. \ref{fig:screening}(a) we plot the exchange interaction energy as a function of screening length for ground ($11$) and excited ($22,33,44$) excitonic $s$ states. Here, we choose the realistically achievable values of $r_0$, which can be tuned by the substrate choice. We observe that for the excited states the growth of $r_0$ leads to the decrease of interaction energy, despite the actual increase of exciton radius $\lambda_n$. This effect can be explained by the evidence that the interaction potential (\ref{pot}) itself decreases rapidly, thus overcoming the impact of interexciton interaction enhancement coming from the exciton wavefunction spread. Hence, one may further increase the XX interaction strength by the reduction of $r_0$, which can be reached by the choice of substrate with large dielectric permittivity.

The situation is different for the exchange interaction of ground state excitons. Fig. \ref{fig:screening} (a) indicates that in the plotted range of screening length the interaction strength varies weakly. This can be seen as a consequence of the non-hydrogenic nature of $1$s excitons in TMDs, and explained as a mutual compensation of interaction enhancement from exciton radius growth and intracarrier interaction decrease. For better understanding we explore the limit $r_0 \rightarrow 0$, where the interaction potential (\ref{pot}) reduces to conventional 2D Coulomb form [\onlinecite{Cudazzo2011}]. In Fig. \ref{fig:screening} (b) we plot the interaction of ground state excitons as a function of screening length in logarithmic scale. We observe that in the above mentioned limit the interaction rapidly decreases, and becomes about $1.3$ times smaller than in screened interaction limit (large $r_0$). Notably, this value is in a close agreement with the previously presented estimate of the ratio (\ref{ratio}) between interactions of excitons in QW and TMD.

We would like to remark also, that in the limit $r\rightarrow 0$ the interaction of excited excitons do not undergo rapid changes, and continues smooth increase (not shown). Such a striking difference of screening length dependence of interaction for ground state and excited states is a direct consequence of the fact, that ground excitonic state in the TMD materials is essentially non-Rydbergian, while excited states demonstrate Rydberg-like behavior [\onlinecite{ChernikovPRL}].

\section{Exciton-electron scattering}

In this section we consider n-doped TMD monolayer with excess of the free electrons which can interact with optically created excitons. This nonlinear process is especially relevant for up to date TMD experiments [\onlinecite{Heinz}], can contribute to the exciton line broadening [\onlinecite{Selig}], and determines the physics of TMD exciton-polarons [\onlinecite{Sidler2017}].
%%%
\begin{figure}
    \includegraphics[width=1.0\linewidth]{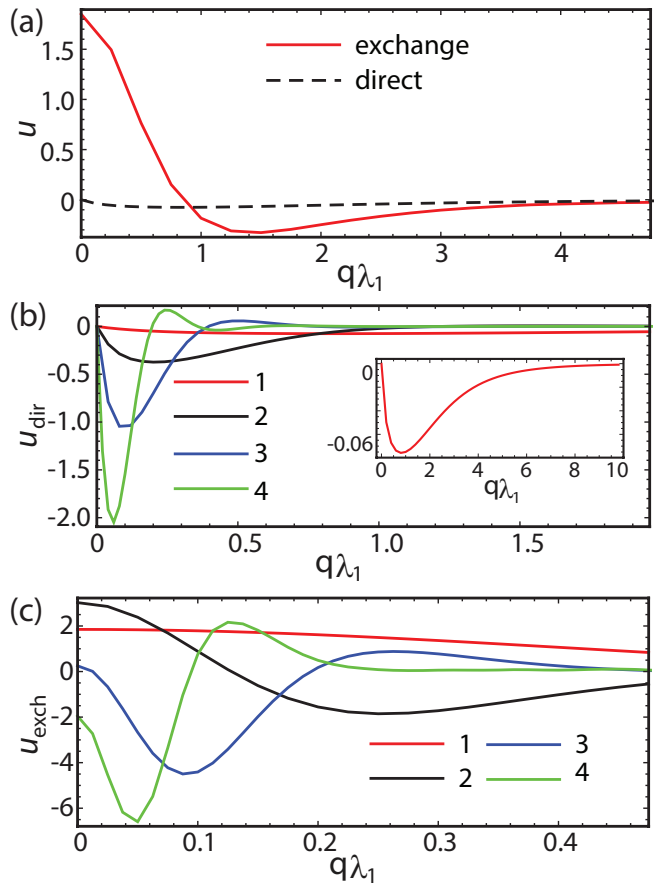}
    \caption{Exciton electron interaction energy as a function of transferred momentum. Ground state direct and exchange interactions (a), direct (b), and exchange(c) interaction of the excited states.}
    \label{e_exc}
\end{figure}
%%%
 We proceed with the calculation of the exciton scattering with conduction band electrons. We restrict our consideration to $s$ states, noting that for $p$-type excitons the results are expected to be similar. The conduction band electron wave function is given by a plane wave $f_{\mathbf{K}}(\mathbf{\rho})=(1/\sqrt{A}) e^{i \mathbf{K} \mathbf{\rho}}$, where $\mathbf{K}$ denotes an electron momentum.  We consider the process of Coulomb scattering of an exciton with an electron, corresponding to the momentum transfer process
\begin{equation}
(n,\mathbf{Q})+(\mathbf{K}) \rightarrow  (n,\mathbf{Q}+\mathbf{q})+(\mathbf{K}-\mathbf{q}).
\end{equation}
Possible interaction channels include direct interaction and the electron exchange term, as shown in Fig. \ref{fig:scheme}(e,f). Correspondingly, one can present the total interaction as sum of the direct and electron exchange contributions:
\begin{align}
&U(n,\mathbf{Q},\mathbf{K},\mathbf{q}) = U_{\mathrm{dir}}(n,q)+U_{\mathrm{exch}}(n,\textbf{q}, \mathbf{K}-\beta_e\mathbf{Q}) \notag \\
       &= \frac{e^2}{4\pi \varepsilon_0} \frac{\lambda_1}{A} \left[u_{\mathrm{dir}}(n,q\lambda_1)+u_{\mathrm{exch}}(n,q\lambda_1, (K-\beta_e Q)\lambda_1)\right],
\end{align}
where the explicit form of the corresponding terms is given in Appendix B. It should be noted that the described approach is in agreement with the method previously used to characterize the exciton-electron scattering in quantum well heterostructures [\onlinecite{Ramon}]. We calculated the scattering of free electron with ground and excited state TMD excitons. Without the loss of generality, it is convenient to put the condition $\mathbf{K}-\beta_e\mathbf{Q}=0$. Fig. \ref{e_exc}(a) illustrates the direct and exchange terms of $1s$ exciton scattering with an electron. Similarly to QW heterostructure, the interaction is governed by exchange contribution. Fig. \ref{e_exc}(b,c) shows direct and exchange interaction of excited excitons with electron, respectively. One can see that similarly to the exciton-exciton interaction both components of scattering amplitudes increase with principal quantum number, conserving the domination of the exchange component. An additional feature is that unlike for the ground state, for the excited states the interaction is attractive and has maxima appearing at intermediate exchange momenta. Moreover, with the increase of quantum number both interaction components become more peak shaped.

\section{Conclusion}

In conclusion, we considered theoretically the exciton-exciton and exciton-electron scattering processes in transition metal dichalcogenide monolayers. We found that unusual screening of the Coulomb interaction characteristic to TMD monolayers leads to the non-monotonic dependence of the exchange interaction on the transferred momentum. We have shown that contrary to the conventional quantum well excitons the interaction can be accurately estimated by a simple analytical formula. It is proportional to the product of the exciton binding energy and the square of exciton radius, and exhibits linear growth with the principal quantum number of exciton. We have studied the dependence of interaction on the dielectric permittivity of a substrate, and have shown that while for excited exciton states interaction increases for the samples with high dielectric permittivity substrates, the ground state interaction cannot be enhanced.

Additionally, we calculated the exciton-electron interaction in TMD monolayers, relevant for systems with excess of free electrons. This interaction is characterized by dominant attractive contribution of the exchange component increasing with the principal quantum number of exciton. The results provide the basis for quantitative description for nonlinear effects in TMD systems, and are important for the design of corresponding nonlinear optoelectronic devices.

\section*{Acknowledgments}

The authors are grateful to D. Gulevich and M. Glazov for fruitful discussions. This work was supported by Icelandic Research Fund, Grant No. 163082-051 and mega-grant No. 14.Y26.31.0015 of the Ministry of Education and Science of Russian Federation.  O.K. thanks University of Iceland for the hospitality during the work on the project, and acknowledges the funding by FP7 ERC Grant QIOS (Grant No. 306576). V.S and I.A.S acknowledge support from Ministry of Education and Science of Russian Federation, goszadanie no 3.2614.2017/4.6 and Horizon2020 RISE project CoExAN.

\appendix
\section{Derivation of matrix elements for Coulomb scattering of Rydberg excitons}

A two-dimensional exciton in $nl$ state with the center-of-mass wave vector $\mathbf{Q}$ is described by the wavefunctions given by Eqs. (\ref{psi_n}) and (\ref{Psi}) in the main text, corresponding to internal and center of mass dynamics, respectively. Considering the states with parallel spin only, one may construct two exciton wave function in the form
\begin{widetext}
\begin{align}
\Phi_{\mathbf{Q},\mathbf{Q}',n}(\mathbf{r}_e,\mathbf{r}_h,\mathbf{r}_{e'},\mathbf{r}_{h'})
&= \frac{1}{2}\left[ \Psi_{\mathbf{Q},n}(\mathbf{r}_e,\mathbf{r}_h) \Psi_{\mathbf{Q}',n}(\mathbf{r}_{e'},\mathbf{r}_{h'}) + \Psi_{\mathbf{Q},n}(\mathbf{r}_{e'}, \mathbf{r}_{h'})\Psi_{\mathbf{Q}', n}(\mathbf{r}_e,\mathbf{r}_h) \right] \notag \\
&- \frac{1}{2}\left[ \Psi_{\mathbf{Q},n}(\mathbf{r}_{e'},\mathbf{r}_h)\Psi_{\mathbf{Q}',n}(\mathbf{r}_e,\mathbf{r}_{h'})+ \Psi_{\mathbf{Q},n}(\mathbf{r}_e,\mathbf{r}_{h'})\Psi_{\mathbf{Q}',n}(\mathbf{r}_{e'},\mathbf{r}_h) \right],
\end{align}
where we omit the magnteic quantum number $m$ for the sake of shortness. The exciton-exciton interaction Hamiltonian reads as
\begin{align}
V_{\mathrm{int}}(\mathbf{r}_e,\mathbf{r}_h,\mathbf{r}_{e'},\mathbf{r}_{h'})= &- V(|\mathbf{r}_e-\mathbf{r}_{h'}|)-V(|\mathbf{r}_{e'}-\mathbf{r}_h|)  \notag \\
                                                                    &+ V(|\mathbf{r}_e-\mathbf{r}_{e'}|)+V(|\mathbf{r}_h-\mathbf{r}_{h'}|),
\end{align}
where all possible interparticle interactions are accounted. The scattering amplitude of the process described by Eq.~(\ref{scattering}) in main text is given by the matrix element
\begin{align}
V_{n}(\mathbf{Q},\mathbf{Q}',\mathbf{q}) &= \int d^2 \mathbf{r}_e d^2 \mathbf{r}_h d^2 \mathbf{r}_{e'} d^2 \mathbf{r}_{h'} \Phi_{\mathbf{Q},\mathbf{Q}',n}^{*}(\mathbf{r}_e,\mathbf{r}_h,\mathbf{r}_{e'},\mathbf{r}_{h'}) V_{int}(\mathbf{r}_e,\mathbf{r}_h,\mathbf{r}_{e'},\mathbf{r}_{h'}) \Phi_{\mathbf{Q}+\mathbf{q},\mathbf{Q}'-\mathbf{q},n}(\mathbf{r}_e,\mathbf{r}_h,\mathbf{r}_{e'},\mathbf{r}_{h'}) \notag \\
&= \int d^2\mathbf{r}_ed^2 \mathbf{r}_hd^2\mathbf{r}_{e'}d^2\mathbf{r}_{h'}\Psi^*_{\mathbf{Q},n}(\mathbf{r}_e,\mathbf{r}_h)\Psi^*_{\mathbf{Q}',n}(\mathbf{r}_{e'},\mathbf{r}_{h'}) V_{int}(\mathbf{r}_e,\mathbf{r}_h,\mathbf{r}_{e'},\mathbf{r}_{h'}) \Psi_{\mathbf{Q}+\mathbf{q},n}(\mathbf{r}_e,\mathbf{r}_h)\Psi_{\mathbf{Q}'-\mathbf{q},n}(\mathbf{r}_{e'},\mathbf{r}_{h'}) \notag \\
&+ \int d^2\mathbf{r}_ed^2 \mathbf{r}_hd^2\mathbf{r}_{e'}d^2\mathbf{r}_{h'}\Psi^*_{\mathbf{Q},n}(\mathbf{r}_e,\mathbf{r}_h)\Psi^*_{\mathbf{Q}',n}(\mathbf{r}_{e'},\mathbf{r}_{h'}) V_{int}(\mathbf{r}_e,\mathbf{r}_h,\mathbf{r}_{e'},\mathbf{r}_{h'}) \Psi_{\mathbf{Q}+\mathbf{q},n}(\mathbf{r}_{e'},\mathbf{r}_{h'})\Psi_{\mathbf{Q}'-\mathbf{q},n}(\mathbf{r}_e,\mathbf{r}_h) \notag \\
&- \int d^2\mathbf{r}_ed^2 \mathbf{r}_hd^2\mathbf{r}_{e'}d^2\mathbf{r}_{h'}\Psi^*_{\mathbf{Q},n}(\mathbf{r}_e,\mathbf{r}_h)\Psi^*_{\mathbf{Q}',n}(\mathbf{r}_{e'},\mathbf{r}_{h'}) V_{int}(\mathbf{r}_e,\mathbf{r}_h,\mathbf{r}_{e'},\mathbf{r}_{h'}) \Psi_{\mathbf{Q}+\mathbf{q},n}(\mathbf{r}_{e'},\mathbf{r}_h)\Psi_{\mathbf{Q}'-\mathbf{q},n}(\mathbf{r}_e,\mathbf{r}_{h'}) \notag \\
&- \int d^2\mathbf{r}_ed^2 \mathbf{r}_hd^2\mathbf{r}_{e'}d^2\mathbf{r}_{h'}\Psi^*_{\mathbf{Q},n}(\mathbf{r}_e,\mathbf{r}_h)\Psi^*_{\mathbf{Q}',n}(\mathbf{r}_{e'},\mathbf{r}_{h'}) V_{int}(\mathbf{r}_e,\mathbf{r}_h,\mathbf{r}_{e'},\mathbf{r}_{h'}) \Psi_{\mathbf{Q}+\mathbf{q},n}(\mathbf{r}_e,\mathbf{r}_{h'})\Psi_{\mathbf{Q}'-\mathbf{q},n}(\mathbf{r}_{e'},\mathbf{r}_h) \notag \\
&= V_{\mathrm{dir}}(n,\mathbf{Q},\mathbf{Q}',\mathbf{q}) +V^X_{\mathrm{exch}}(n,\mathbf{Q},\mathbf{Q}',\mathbf{q}) +V^e_{\mathrm{exch}}(n,\mathbf{Q},\mathbf{Q}',\mathbf{q})+V^h_{\mathrm{exch}}(n,\mathbf{Q},\mathbf{Q}',\mathbf{q}), \notag \\
\end{align}
where four terms correspond to direct interaction, exciton exchange, electron exchange, and hole exchange. Before proceeding further, one can note that for the case when $\mathbf{Q}=\mathbf{Q}'$, we have
\begin{align}
&V^X_{\mathrm{exch}}(n,\mathbf{Q},\mathbf{Q},\mathbf{q}) = V_{\mathrm{dir}}(n,\mathbf{Q},\mathbf{Q},\mathbf{q}),    \notag \\
&V^h_{\mathrm{exch}}(n,\mathbf{Q},\mathbf{Q},\mathbf{q}) = V^e_{\mathrm{exch}}(n,\mathbf{Q},\mathbf{Q},\mathbf{q}).
\end{align}
Introducing dimensionless functions $\widetilde{V}(x)$ and $\widetilde{\psi}_n(x)$ as
\begin{equation}
V(r)=- \frac{e_1 e_2}{4 \pi \varepsilon_0 \lambda_n} \frac{1}{r_0/\lambda_n}\left[\mathrm{ln}\left(\frac{r/\lambda_n}{r/\lambda_n+r_0/\lambda_n}\right)-(\gamma-\mathrm{ln}2)e^{-\frac{r/\lambda_n}{r_0/\lambda_n}} \right] := \frac{e_1 e_2}{4 \pi \varepsilon_0 \lambda_n} \widetilde{V}\left(\frac{r}{\lambda_n}\right),
\end{equation}
\begin{equation}
\psi(r) := \frac{1}{\lambda_n} \widetilde{\psi}_n\left(\frac{r}{\lambda_n}\right),
\end{equation}
one may write the direct term in the explicit form as
\begin{align}
V_{\mathrm{dir}}(n,q) = \frac{e^2}{4 \pi \varepsilon_0} \frac{\lambda_1}{A} I_{\mathrm{dir}} (n,q) &:= \frac{e^2}{4 \pi \varepsilon_0} \frac{\lambda_1} {A}  \lambda \int d^2 \mathbf{x} d^2 \mathbf{x}' d^2 \mathbf{\xi} e^{i \mathbf{q}\lambda_1 \lambda \mathbf{\xi} } \widetilde{\psi}_n^2(x) \widetilde{\psi}_n^2(x') \left[ -\widetilde{V}(|\mathbf{\xi}+\beta_h\mathbf{x}+\beta_e\mathbf{x}'|) \right. \notag \\
 &- \left.  \widetilde{V}(|\mathbf{\xi}-\beta_e\mathbf{x}-\beta_h\mathbf{x}'|) +\widetilde{V}(|\mathbf{\xi}+\beta_h(\mathbf{x}-\mathbf{x}')|) +\widetilde{V}(|\mathbf{\xi}-\beta_e(\mathbf{x}-\mathbf{x}')|) \right],
\end{align}
where we have introduced the notations $\mathbf{r}=\mathbf{r}_e-\mathbf{r}_h$, $\mathbf{R}=\beta_e\mathbf{r}_e+\beta_h\mathbf{r}_h$, $\mathbf{r}'=\mathbf{r}_{e'}-\mathbf{r}_{h'}$, $\mathbf{R}'=\beta_e\mathbf{r}_{e'}+\beta_h\mathbf{r}_{h'}$, $\mathbf{\xi}=\frac{\mathbf{R}-\mathbf{R}'}{\lambda_n}$,  $\mathbf{x}=\frac{\mathbf{r}}{\lambda_n}$, $\mathbf{x}'=\frac{\mathbf{r}'}{\lambda_n}$, $\lambda=\frac{\lambda_n}{\lambda_1}$. The remaining step consists in performing integrations. Considering the first term, we can rewrite integral as
\begin{align}
&\int d^2 \mathbf{x} d^2 \mathbf{x}' d^2 \mathbf{\xi} e^{i \mathbf{q}\lambda_1 \lambda \mathbf{\xi} } \widetilde{\psi}_n^2(x) \widetilde{\psi}_n^2(x')  \widetilde{V}(|\mathbf{\xi}+\beta_h\mathbf{x}+\beta_e\mathbf{x}'|) = \int d^2 \mathbf{\tau} e^{i \mathbf{q}\lambda_1 \lambda \mathbf{\tau} } \widetilde{V}(\tau) \int d^2 \mathbf{x}  e^{-i \mathbf{q}\lambda_1 \lambda \beta_h \mathbf{x} } \widetilde{\psi}_n^2(x) \int d^2 \mathbf{x}' e^{-i \mathbf{q}\lambda_1 \lambda \beta_e \mathbf{x}' } \widetilde{\psi}_n^2(x') \notag \\
&= (2\pi)^3 \int\limits_0^\infty J_0 (q \lambda_1 \lambda \tau) \widetilde{V}(\tau) \tau d\tau \int\limits_0^\infty J_0 (q \lambda_1 \lambda \beta_h x) \widetilde{\psi}_n(x) x d x \int\limits_0^\infty J_0 (q \lambda_1 \lambda \beta_e x') \widetilde{\psi}_n(x') x' d x' = (2\pi)^3 V_{q \lambda_1 \lambda} g_n(\lambda \beta_h q \lambda_1) g_n(\lambda \beta_e q \lambda_1),
\end{align}
where the functions are defined as
\begin{equation}
  V_{q}=\int\limits_0^\infty J_0 (q \tau) \widetilde{V}(\tau) \tau d\tau,
\end{equation}
\begin{equation}
  g_{n}(q)=\int\limits_0^\infty J_0 (q  x) \widetilde{\psi}_n(x) x d x.
\end{equation}
Calculating the remaining terms in the same way, we arrive to
\begin{equation}
I_{\mathrm{dir}} (n,q) = \lambda (2\pi)^3 V_{\lambda_1 q \lambda } \left[g_n(\beta_h \lambda \lambda_1 q)- g_n(\beta_e \lambda \lambda_1 q)\right]^2.
\end{equation}
The electron exchange integral after some simplifications takes a form
\begin{align}
V^e_{\mathrm{exch}}(n,q)=\frac{e^2}{4 \pi \varepsilon_0} \frac{\lambda_1}{A} I^e_{\mathrm{exch}} (n,q) = \frac{e^2}{4 \pi \varepsilon_0} \frac{\lambda_1} {A}  \lambda & \int & d^2 \mathbf{x} d^2 \mathbf{y}_1 d^2 \mathbf{y}_2 e^{i \mathbf{q}\lambda_1 \lambda (\beta_h\mathbf{y}_1-\beta_e\mathbf{y}_2-\mathbf{x}) } \widetilde{\psi}_n(x) \widetilde{\psi}_n(y_1) \widetilde{\psi}_n(y_2) \widetilde{\psi}_n(|\mathbf{y}_2-\mathbf{y}_1-\mathbf{x}|) \notag \\
& & \left[ \widetilde{V}(y_1)+\widetilde{V}(y_2)-\widetilde{V}(|\mathbf{y}_1+\mathbf{x}|)-\widetilde{V}(|\mathbf{y}_2-\mathbf{x}|)  \right].
\end{align}

\section{Derivation of matrix elements for the Coulomb scattering of Rydberg excitons with electrons}

In the following section we demonstrate the derivation of the exciton-electron interaction matrix elements. The wave function of an exciton-electron pair can be written in the form
\begin{equation}
F_{\mathbf{Q},\mathbf{K},n}(\mathbf{r}_1,\mathbf{r}_2,\mathbf{r}_h) =\frac{1}{\sqrt{2}} \left[ \psi_{\mathbf{Q},n}(\mathbf{r}_1,\mathbf{r}_h) f_{\mathbf{K}}(\mathbf{r}_2) -\psi_{\mathbf{Q},n}(\mathbf{r}_2,\mathbf{r}_h) f_{\mathbf{K}}(\mathbf{r}_1) \right]
\end{equation}
The interaction Hamiltonian is given by
\begin{equation}
\hat{U}_{\mathrm{int}} = -V(|\mathbf{r}_1-\mathbf{r}_h|)-V(|\mathbf{r}_2-\mathbf{r}_h|) +V(|\mathbf{r}_1-\mathbf{r}_2|).
\end{equation}
The full scattering matrix element reads as
\begin{align}
U_{n}^{\mathrm{full}}(\mathbf{Q},\mathbf{K},\mathbf{q}) &= \int d^2 \mathbf{r}_1 d^2 \mathbf{r}_2 d^2 \mathbf{r}_h F_{\mathbf{Q},\mathbf{K},n}^{*}(\mathbf{r}_1,\mathbf{r}_2,\mathbf{r}_h) \hat{U}_{int}(\mathbf{r}_1,\mathbf{r}_2,\mathbf{r}_h) F_{\mathbf{Q}+\mathbf{q},\mathbf{K}-\mathbf{q},n}(\mathbf{r}_1,\mathbf{r}_2,\mathbf{r}_h)  \notag \\
&= \frac{1}{2A^2} \int d^2 \mathbf{r}_1 d^2 \mathbf{r}_2 d^2 \mathbf{r}_h \hat{U}_{int}(\mathbf{r}_1,\mathbf{r}_2,\mathbf{r}_h)
\left\{ e^{i\mathbf{q}(\beta_e\mathbf{r}_1+\beta_h\mathbf{r}_h-\mathbf{r}_2)} \psi_n^2 (\mathbf{r}_1-\mathbf{r}_h) +e^{i\mathbf{q}(\beta_e\mathbf{r}_2+\beta_h\mathbf{r}_h-\mathbf{r}_1)} \psi_n^2 (\mathbf{r}_2-\mathbf{r}_h) \right.  \notag \\
&   \left. -\left[ e^{i (\alpha\mathbf{Q}-\mathbf{K})(\mathbf{r}_2-\mathbf{r}_1)} e^{i \mathbf{q}(\beta_e\mathbf{r}_1+\beta_h\mathbf{r}_h-\mathbf{r}_2)} +e^{-i (\alpha\mathbf{Q}-\mathbf{K})(\mathbf{r}_2-\mathbf{r}_1)} e^{i\mathbf{q}(\beta_e\mathbf{r}_2+\beta_h\mathbf{r}_h-\mathbf{r}_1)} \right] \psi_n (\mathbf{r}_1-\mathbf{r}_h) \psi_n (\mathbf{r}_2-\mathbf{r}_h)  \right\}. \notag \\
\end{align}
Here we note that the above expression contains terms contributing to the exciton internal dynamics. For instance, the term $-V(|\mathbf{r}_1-\mathbf{r}_h|) e^{i\mathbf{q}(\beta_e\mathbf{r}_1+\beta_h\mathbf{r}_h-\mathbf{r}_2)} \psi_n^2 (\mathbf{r}_1-\mathbf{r}_h)$ describes the interaction between hole and electron $\mathbf{r}_1$, forming exciton, while the second electron is not involved in the system. Hence, this term should be neglected. Analogously, excluding all the extra terms and after corresponding simplifications, one finally arrives to the sum of direct and exchange components
\begin{equation}
U_{n}(\mathbf{Q},\mathbf{K},\mathbf{q})=U_{\mathrm{dir}}(n,q)+U_{\mathrm{exch}}(n,\textbf{q}, \mathbf{K}-\beta_e\mathbf{Q}) = \frac{e^2}{4\pi \varepsilon_0} \frac{\lambda_1}{A} \left[u_{\mathrm{dir}}(n,q\lambda_1)+u_{\mathrm{exch}}(n,q\lambda_1, (K-\beta_eQ)\lambda_1)\right],
\end{equation}
where
\begin{equation}
u_{\mathrm{dir}} (n,q) =\lambda (2\pi)^2 V_{q \lambda_1 \lambda} \left[g_n(\lambda \beta_h q \lambda_1)- g_n(\lambda \beta_e q \lambda_1)\right],
\end{equation}
and
\begin{align}
u_{\mathrm{exch}}(n,q)= \lambda \int d^2 \mathbf{\chi} d^2 \mathbf{x} e^{-i \mathbf{q}\lambda_1 \lambda (\alpha\mathbf{\chi}+\mathbf{x})}
e^{i (\mathbf{K}-\alpha\mathbf{Q})\lambda_1 \lambda (\mathbf{\chi}+\mathbf{x})}
\left[ -\widetilde{V}(\chi) +\widetilde{V}(|\mathbf{\chi}+\mathbf{x}|) \right] \widetilde{\psi}_n(\chi) \widetilde{\psi}_n(x),
\end{align}
with $\mathbf{\chi}=\frac{\mathbf{R}-\mathbf{r}_2-\beta_e\mathbf{r}}{\lambda_n}$.

\end{widetext}

%%%%%%%%%%%%%%%%%%%%%%%%%%%%%%%%%

\end{document}